# False Reassurance in LLM Severe-Weather Guidance: A Replayable Evaluation Against Real NWS Alerts

Kadharmoideen Fadurudeen

*Independent Researcher (e-mail: kadhar.mfn@gmail.com; ORCID: 0009-0002-4526-8812)*

***Abstract*—An LLM assistant that cannot see official severe-weather alerts may tell users that conditions are calm while a warning is active. We call this false reassurance and measure it by replaying 117 frozen cases built from real U.S. National Weather Service (NWS) alerts (69 alert cases, 32 no-alert controls, 16 non-U.S. locations) through five conditions of a Claude-based preparedness pipeline that differ in alert access and prompt design. One rater hand-labeled all 342 alert-case responses under a seven-category scheme written after an initial binary pass, which we disclose. False reassurance was 0.0% for the production pipeline (0/66, 95% CI 0.0–5.5) and for the same alerts under a neutral prompt (0/69), 37.7% for a blind neutral prompt (26/69), and 44.9% when the blind prompt also told the model to disclose its blind spot (31/69). Every case was implicit; no response explicitly denied an alert. Real alerts removed false reassurance (paired McNemar $p < 0.0001$); the disclosure instruction did not ($p = 0.40$). An automated regex measure suggested the opposite and missed 53 of 57 true cases, so we do not use it. Alert data alone was not enough for safe behavior: with a neutral prompt the model raised some alarm on 87.5% of no-alert controls, against 0% under the production prompt, and a preregistered single-sentence ablation showed that one production instruction explains part of that gap. We also report two failures common to similar systems, NWS test products ingested as real alerts and a token budget that fails on complex cases, and we release the dataset, labels and code.**

## I. INTRODUCTION

Voice assistants and chat assistants are being asked what to do before and during severe weather. A wrong answer is costly in a specific way: an answer that sounds calm during an active warning can delay protective action, while one that raises alarm on a quiet day teaches users to ignore the assistant. Official alerts exist so that people do not have to guess, but a language model only uses them if the system around it fetches them and the prompt tells the model how to treat them.

This paper grew out of one observation. While testing a preparedness assistant in Tucson, Arizona during a live Flash Flood Warning, we compared the full pipeline with a baseline that saw only season and current conditions. The baseline did not merely give vaguer advice. It stated that there were no active alerts, when the National Weather Service (NWS) had issued two. That is a factual error about the world, not a quality gap, and a single anecdote cannot say how common it is. We therefore turned the incident into a measurement.

We ask three questions:

- **RQ1.** How often does an LLM guidance assistant without alert access tell users that no alert or hazard exists when an official alert is active?
- **RQ2.** How much does live official alert data reduce this, and does telling the model to disclose its blind spot help?
- **RQ3.** Does supplying alerts guarantee appropriate behavior, including on days with no alert?

To answer them we log real NWS alerts as they happen, freeze them into replayable cases, and run the same cases through controlled variants of one pipeline. Replay matters because NWS serves only active alerts, so a case that is not recorded when it occurs cannot be reconstructed.

**Contributions.**

1. A logging and case-construction method for real official alerts, and a frozen set of 117 replayable cases (69 alert cases, 32 no-alert controls, 16 non-U.S. cases) that we plan to release.
2. A measurement of false reassurance and its mirror image, false alarm, with explicit and implicit forms reported separately.
3. A five-condition design, plus a preregistered two-condition single-sentence ablation, separating the effect of alert data from the effect of prompt and schema design, with evidence that data access alone did not yield safe behavior on no-alert cases.
4. Two practical failure findings: NWS test products ingested as real alerts, and a token budget that fails on the most complex alert cases.

This is a pilot-scale study of one pipeline and one model family in one season, and we say so throughout. Sections II to VI describe related work, the system, the data, the design and the results. Sections VII and VIII discuss implications and limits.

## II. RELATED WORK

**Tool failure and honest reporting.** The closest work is on how tool-using agents report what they could and could not do. Failure-Transparent Agents [1] builds 100 tasks with fixed, deterministic tool failures and checks each claim the agent makes against the evidence it had. It reports false-success rates falling from 22.8% at baseline to 9.3% with a transparency instruction and 0.8% with a structured evidence contract. ToolFailBench [2] diagnoses tool-use failures across 1,000 tasks in five domains and pairs tool-required tasks with control tasks that need no tool. Our setting differs in three ways. The failure is silent absence: the model is not shown an error, it simply never receives the alert, and cannot tell that something is missing. The ground truth is real, time-stamped official alerts rather than injected failures. And the harm of a wrong answer runs in both directions, so we measure false alarms on no-alert controls as well as false reassurance. Our finding that a disclosure instruction does not reliably reduce false reassurance

— it was statistically indistinguishable from no disclosure, and numerically higher (Section VI) — contrasts with the transparency-instruction effect reported in [1]; we do not have an explanation for the difference beyond the three setting differences just described.

**LLMs for disasters and emergencies.** DisasterQA [3] tests LLMs on disaster-response questions drawn from online sources and measures answer accuracy and confidence. A recent survey [4] organizes LLM work on disaster management by phase and application. SafeMate [5] is a retrieval-based emergency assistant built on the Model Context Protocol that retrieves from trusted documents. In the sources we reviewed, we did not find an evaluation that replays real, time-stamped official alerts and scores whether an assistant wrongly denies that they exist. We read these works through their abstracts and may have missed relevant evaluations.

**Hallucination and grounding.** Surveys of hallucination in text generation [6] describe outputs that are fluent but unsupported by the source or the world. Retrieval and tool use are the standard remedy, giving the model access to current facts at generation time [7]. The Tucson incident is a case where that remedy was absent, and the model filled the gap with a confident claim. Our question is how much of the remedy comes from the data and how much from the instructions about using it.

**False alarms in public warning.** Social science on warnings treats false alarms as a calibration problem with mixed evidence. Trainor et al. [8] found that public definitions of a false alarm vary widely, and that county-level false-alarm ratios did not predict perceptions of false alarms but did predict response. Lim et al. [9], in surveys of 4,162 Southeastern U.S. residents, found that fears of a "cry wolf" effect may be exaggerated. We therefore treat false alarm as a measurable calibration error of the assistant, and make no claim about how users respond to it.

**Data sources.** The NWS web API serves currently active alerts and recent history through an open interface [10], and Open-Meteo provides current conditions under a Creative Commons licence [11]. Because the NWS interface serves active alerts, historical states must be logged as they happen, which motivates our replay design.

## III. SYSTEM UNDER TEST

We study Stormwise, an open-source Model Context Protocol (MCP) server that gives a voice assistant severe-weather preparedness guidance. A request for a location goes through four steps:

1. The location is geocoded to coordinates.
2. Current conditions are fetched from Open-Meteo, and active alerts for that point are fetched from the NWS `alerts/active` endpoint, in parallel.
3. A Claude model on Amazon Bedrock is called with forced tool use, so it must return a structured `prep_guidance` result: a `primary_concern`, a `severity` (none, minor, moderate, severe or extreme), a list of `prep_steps`, and its `reasoning`.
4. Locations outside NWS coverage are marked "unsupported" and the model is told so, rather than being given an empty alert list.

Two design details matter for the experiments. First, the production pipeline passes the model only each alert's event name, severity, urgency, affected area and headline. The full alert description and the official instructions are not passed, so "alert-aware" in this paper means headline-level awareness. Second, the production system prompt includes an explicit sentence for the empty case: if there are no active alerts, say so plainly and briefly, and do not invent urgency. The production tool schema also contained the example phrase "no active alerts right now" in the description of `primary_concern`. We removed that phrase in our neutral conditions, because it is a plausible confound for any false-reassurance measurement (Section V).

The model was called at temperature 0 with a 500-token output limit, the production setting. The main study ran on 2026-10-23 (17:47 to 18:46 UTC) using Claude Sonnet 4.6 via an Amazon Bedrock inference profile. All 585 study calls used this single model, so results describe one model family and should not be read as a ranking of models.

## IV. DATASET

**Collection.** A logger written in Python's standard library polled 67 locations: 61 U.S. cities spread across nine regional groups chosen to cover different hazard climates, and 6 non-U.S. cities that NWS does not cover. Every 10 minutes it requested active alerts for each point and stored the full alert properties. Hourly it stored Open-Meteo conditions. It also polled the nationwide active-alert feed every 10 minutes and de-duplicated alerts by identifier. The collection window ran from 2026-10-19 to 2026-10-23 (96.6 hours). City coordinates are approximate city centres, not exact addresses.

**Gaps.** The collection machine slept for 74.2 of the 96.6 hours (77%). Alerts that began and ended inside a gap were never recorded, so short-lived warnings such as flash flood and tornado warnings are the most likely to be missing. The dataset is therefore biased toward longer-lived alerts and advisories, and the bias cannot be repaired afterwards.

**Case construction.** A build script selected cases by stratified sampling over hazard family and severity with a fixed seed (42), and reported shortfalls rather than filling them. Hazard family was assigned from the NWS event name by fixed rules. The frozen set has 117 cases:

| Group | Cases | Purpose |
|---|---|---|
| Alert cases (7 hazard families) | 69 | Measure false reassurance, hazard match, severity |
| No-alert controls (U.S. locations, no active alert) | 32 | Measure false alarm |
| Unsupported locations (non-U.S.) | 16 | Measure honesty about missing coverage |

The earlier build contained 79 alert cases before the test-product fix described below. Table 1 shows the frozen 69 alert cases by hazard family and NWS severity, together with the number of independent events in each family.

**Table 1. Alert cases by hazard family and NWS severity.**

| Hazard family | Minor | Moderate | Severe | Unknown | Cases | Independent events |
|---|---|---|---|---|---|---|
| Coastal | – | 21 | – | – | 21 | 21 |
| Flood | 9 | – | 4 | – | 13 | 13 |
| Other | – | 10 | – | – | 10 | 9 |
| Severe thunderstorm | – | – | 5 | – | 5 | 3 |
| Winter | 5 | 1 | – | – | 6 | 6 |
| Fog | – | 3 | – | – | 3 | 2 |
| Air quality | – | – | – | 11 | 11 | 7 |
| Total | 14 | 35 | 9 | 11 | 69 | 61 |

Independent events are counted within each hazard family. Two event clusters span two families: a Los Angeles case classified as "other" shares an alert with a severe-thunderstorm case, and a Phoenix flood case shares an alert with a severe-thunderstorm case. The per-family counts therefore sum to 61, while counting across the whole set gives 59, which is the figure we use for the overall independent sample size. No alert in the final set has NWS severity Extreme, so the study cannot test behaviour on the most severe alerts. Cells with fewer than 10 cases are descriptive only.

Each case stores the location, the Open-Meteo conditions, the full NWS alert properties, the coverage flag and a timestamp, so it can be replayed without calling any live service.

**Test products.** NWS periodically broadcasts nationwide test products with status "Test". An earlier build of our dataset contained a whole "tsunami" hazard family (10 of 10 cases) that turned out to be test warnings, not real ones. We found this by reading a response in which the model correctly recognised a test alert as not applicable. We now exclude every alert whose status is not "Actual" and we removed the tsunami family. The production client we studied does not filter on status either, so it can present a test product to a user as a real alert.

**Independent events.** The 69 alert cases refer to 85 distinct alert identifiers, but only 59 independent event clusters once the same alert, re-logged as a new case because the set of concurrent alerts changed between polls, is merged. No event was shared across different cities. We report 59 as the independent sample size for any analysis sensitive to repeated measurement of the same event.

## V. EXPERIMENTAL DESIGN

Each frozen case was replayed through five conditions without calling any live service. The conditions differ in two things: whether the model sees the real alerts, and which prompt and schema it uses.

| Condition | Sees real alerts | Prompt and schema | Purpose |
|---|---|---|---|
| A, full | Yes (headline-level) | Production prompt and schema | The system as shipped |
| A2, alerts only | Yes (headline-level) | Neutral prompt and schema, as in B | Isolates alert data from prompt wording |
| B, blind neutral | No | Neutral: the "no active alerts right now" example removed, and no production sentence on the empty case | Baseline without the schema confound |
| C, blind with disclosure | No | As B, plus an instruction to say it cannot see live alerts and to point to official sources | Tests a disclosure instruction |
| D, plain | No | None: "What should I do to prepare in <location>?" | Bare model reference |

The only difference between A2 and B is the presence of alerts. The only difference between A and A2 is the prompt and schema. Comparing A with B alone would mix the two, which is why A2 was added.

**Preregistered ablation.** After the main study we preregistered one hypothesis and tested it with two further conditions on all 117 cases. A2plus is A2 plus the single production sentence about the empty case and nothing else. Aminus is the production prompt and schema with only that sentence removed. The planned comparisons are A2plus against A2 and Aminus against A. False alarm is also reported at a stricter threshold of severity "moderate" or higher.

**Metrics.** False reassurance is hand-labeled (see Manual validation below). False alarm, hazard match and severity are rule-based, with every match flagged for manual review.

- **False reassurance** (alert cases only): the response claims that no alert exists (explicit) or that no hazard or threat exists (implicit). A response that honestly says it cannot see live alerts is counted separately as disclosure, not as reassurance.
- **False alarm** (no-alert controls only): the model returns a severity other than "none" or, as a stricter reading, "moderate" or higher.
- **Hazard match:** the response names the hazard family of the real alert.
- **Severity exact match:** the model's severity equals the NWS severity on the same ordinal scale.
- **Unsupported-location honesty** (non-U.S. cases): the response says it has no alert coverage, rather than claiming there are no alerts.

Hazard match and severity are reported as "what the alert data adds", not as model accuracy, because the blind conditions cannot know the alert. Condition D has no severity field, so severity and false-alarm metrics are not applicable to it.

**Statistics.** Proportions carry 95% Wilson confidence intervals [12]. Paired comparisons use McNemar's test with continuity correction on the same cases [13]. Breakdown cells with fewer than 10 cases are descriptive only.

**Cost and calls.** The main study made 585 Bedrock calls, of which 581 succeeded, for a total of $4.73. A separate preregistered ablation (A2plus and Aminus, Section VI) added 234 further calls at $2.30, with 0 failures.

**Manual validation.** The author hand-labeled every alert-case response from every condition (66 for condition A, whose three failed calls returned nothing, and 69 for each of the other four conditions, 342 in all) in a shuffled copy with condition labels and automated flags hidden. Labeling proceeded in two stages. A first batch of 108 responses (64 condition-A responses, the 22 responses the regex flagged as false reassurance, and 22 unflagged responses as a check for missed cases; the flagged set includes two more condition-A responses, which completes A's 66) was labeled with a binary label, false reassurance or not. That pass flagged 25 of the 108, including 6 of condition A's 66 (9.1%), and agreed poorly with the regex (28.6% precision, 24.0% recall). Reading the six condition-A cases against the paper's own definition showed that none denied the real alert or hazard: they named it and either minimized it or engaged with it correctly. We then wrote a seven-category scheme (denies the alert, denies the hazard, acknowledges-but-downplays, discloses the blind spot, non-answer, engages correctly, wrong hazard addressed without denial), applied it to all 108 responses, and then to the remaining 234, so that every alert-case response has a category. Only the first two categories count as false reassurance; downplaying and wrong-hazard responses are reported separately (Table 3). The scheme is frozen and will be released with the labels. Two points limit what this validation can show. First, the scheme was written after the binary pass had returned a result for the production pipeline, and applying it moved that result from 9.1% to 0.0%, which is a researcher-degrees-of-freedom risk whatever the merits of the categories. In mitigation, the scheme was applied to all responses rather than only the six in question, it moved responses in both directions (of the 25 first-batch responses the binary pass flagged, 22 were re-categorized as not false reassurance, and 4 responses the binary pass had not flagged were categorized as implicit false reassurance), and the quoted responses can be checked against the released labels. Second, there is one rater, who also built the system under test, and no reliability statistic. The labeling hid condition labels but not response content, which can reveal the condition (a response that cites a specific warning plainly had alerts). Against the final labels, the regex had 19.0% precision (4/21) and 7.0% recall (4/57) over all 342 responses, so we do not use it for any false-reassurance figure. Judge-based scoring of advice correctness and harmful advice is built but was not run, so this paper makes no claim about the quality of the advice itself.

## VI. RESULTS

All false-reassurance figures in this section are hand labels over every alert-case response (Section V). False alarm, hazard match and severity are rule-based. False reassurance is computed over alert cases (66 for condition A, 69 for the others), hazard match and severity over alert cases, and false alarm over the 32 no-alert controls.

**Table 2. Main results by condition.** Percentages with 95% Wilson confidence intervals where available.

| Condition | Alert cases (n) | False reassurance | Explicit | Implicit | False alarm on controls | Hazard match | Severity exact |
|---|---|---|---|---|---|---|---|
| A, full | 66 | 0.0% (0.0–5.5) | 0.0% | 0.0% | 0.0% | 86.4% | 86.7% |
| A2, alerts only | 69 | 0.0% (0.0–5.3) | 0.0% | 0.0% | 87.5% (71.9–95.0) | 85.5% | 53.0% |
| B, blind neutral | 69 | 37.7% (27.2–49.5) | 0.0% | 37.7% | 96.9% | 33.3% | 23.9% |
| C, blind with disclosure | 69 | 44.9% (33.8–56.6) | 0.0% | 44.9% | 100.0% | 30.4% | 23.9% |
| D, plain | 69 | 0.0% (0.0–5.3)† | 0.0% | 0.0% | n/a | 18.8% | n/a |

† Condition D returned a clarifying question about what the user was preparing for in all 69 alert cases and gave no guidance about the location, so its 0.0% is a non-answer, not a calibration result, and it is not compared with the others.

Note. False-reassurance figures for all five conditions are hand labels over every alert-case response (66 for A, 69 for the others), assigned by a single rater (the author) with a seven-category scheme that was written after an initial binary pass (Section V). They count only responses that deny an alert or a hazard; responses that downplay the alert or address a different hazard are counted separately in Table 3. False alarm, hazard match and severity remain rule-based.

Table 3. Category assigned by hand to every alert-case response, by condition (counts). No response was coded as explicitly denying that an alert existed, so the implicit column is all of the false reassurance. Downplaying and wrong-hazard responses are not counted as false reassurance.

| Condition | Implicit false reassurance | Downplays the alert | Wrong hazard, no denial | Engages correctly | Discloses blind spot | Non-answer | Total |
|---|---|---|---|---|---|---|---|
| A, full | 0 | 3 | 4 | 58 | 0 | 1 | 66 |
| A2, alerts only | 0 | 1 | 2 | 65 | 0 | 1 | 69 |
| B, blind neutral | 26 | 7 | 19 | 17 | 0 | 0 | 69 |
| C, blind with disclosure | 31 | 3 | 13 | 9 | 12 | 1 | 69 |
| D, plain | 0 | 0 | 0 | 0 | 0 | 69 | 69 |

**RQ1: false reassurance without alerts.** With no alert data and a neutral prompt (B), 37.7% of alert-case responses (26/69, 95% CI 27.2–49.5) were false reassurance on the hand labels, all of the implicit kind: the response described calm or unremarkable conditions or said there was no immediate threat while an official alert was active. No response in any condition explicitly stated that no alert existed, so the explicit pattern seen in the Tucson incident did not recur in this replay. Another 7 of B's responses downplayed the alert and 19 addressed a different hazard without denying anything (Table 3); we report these separately and do not count them as false reassurance. The automated regex had put B at 21.7%; it undercounted, flagging 15 of B's responses, of which the hand labels confirmed 3. With the production pipeline (A), the hand-labeled rate is 0.0% (0/66, 95% CI 0.0–5.5), not the regex's 3.0% and not the 9.1% of the first binary pass (Section V). A paired McNemar test over the 66 cases that both conditions answered finds 25 cases of false reassurance under B and none under A ($\chi^2 = 23.04$, $p < 0.0001$, continuity-corrected).

**RQ2: effect of alert data and of a disclosure instruction.** Alerts under the neutral prompt (A2) produced no false reassurance in 69 cases, against 26 for B on the same cases (26 discordant cases in one direction and none in the other, $\chi^2 = 24.04$, $p < 0.0001$). A and A2 did not differ (0 of 66 each). Telling the blind model to disclose its blind spot (C) did not reduce false reassurance: C was 44.9% (31/69, 95% CI 33.8–56.6), numerically higher than B, and the difference is not significant (9 cases only B, 14 cases only C; $\chi^2 = 0.70$, $p = 0.40$). The instruction did change what the model said: 12 of C's responses disclosed the blind spot and none of B's did (Table 3), but 31 of C's responses still described calm conditions or no immediate threat. Because the scheme assigns a response that discloses to the disclosure category first, a response that both disclosed and reassured is not counted here, so these rates are conservative. C differs from the production pipeline (31 cases only C, none only A, $\chi^2 = 26.04$, $p < 0.0001$). This is the opposite of what the regex suggested (5.8% for C) and of the transparency-instruction effect reported for tool failures [1], and we did not expect it. Alert data also changed hazard identification and severity (Table 2), which we read as what the alerts add, not as model accuracy.

**RQ3: alerts do not guarantee appropriate behavior.** A2 had the same real alert data as A, yet it returned a non-"none" severity on 87.5% of the 32 no-alert controls, against 0.0% for A, and its exact severity match was 53.0% against 86.7%. The blind conditions were at 96.9% and 100.0%. A preregistered follow-up isolated the one sentence in the production prompt that tells the model how to treat the empty case: adding it alone to A2's neutral prompt and schema (condition A2plus) cut the false-alarm rate from 87.5% to 31.3% (10/32), a highly significant reduction (McNemar $\chi^2 = 16.056$, $p < 0.001$) that did not fully close the gap to A's 0%. Removing the same sentence from the full production prompt (condition Aminus) barely moved its rate (0.0% to 3.1%, 1/32; $\chi^2 = 0.000$, $p = 1.000$), statistically indistinguishable from A. The sentence is therefore enough to do substantial work on its own, though not enough to reach the production prompt's rate, and it is not necessary once everything else in the production prompt is present, so some other part of that prompt or schema is also doing work. Neither ablation condition's false-reassurance rate changed significantly from its baseline (A2plus vs A2 $p = 0.248$; Aminus vs A $p = 0.617$). These two comparisons use the automated regex, because the ablation responses were not hand-labeled, and they cannot separate the sentence's effect from a floor effect: false reassurance was already at or near zero in both baselines, so there was little room to change.

False alarm above means any severity other than "none". At a stricter "moderate or higher" threshold the picture changes sharply: A and D remain at 0.0% (0/32), A2 drops to 3.1% (1/32), while B and C stay high at 46.9% (15/32) each. Severity distributions on the 32 controls: A {none: 32}; A2 {minor: 27, none: 4, moderate: 1}; B {minor: 16, moderate: 15, none: 1}; C {minor: 17, moderate: 15}; D {unparsed: 32, since D has no severity field}. At the permissive "more than none" threshold, A2 looks nearly as unsafe as the blind conditions B and C; at the stricter threshold, which is closer to a real consequence, A2 is far closer to A's 0% than to B/C's roughly 47%.

**Production failures.** Four of the 585 calls failed with "missing required fields," all under the production pipeline (condition A). One was a non-U.S. unsupported-location case (Chennai); the other three were alert cases: one flood case in Phoenix, Arizona, and two cases in Los Angeles, California (one severe-thunderstorm, one unclassified "other"). Replaying the failing cases at the production settings succeeded and used 439 to 488 of the 500-token output budget, and with a limit of 1,200 tokens the same cases used 405 to 498 tokens. These cases therefore need almost the whole budget, and ordinary run-to-run variation at temperature 0 is enough to tip them over. This explanation is inferred from the replays, not observed on the failing calls. Excluding the three failed alert cases leaves 66 hand-labeled responses with no false reassurance (0.0%, 95% CI 0.0–5.5); counting each failure as a non-answer ("no guidance returned") instead gives 0.0% false reassurance (0/69) and a 4.3% no-guidance rate (3/69), so excluding the failures does not flatter condition A on this measure, although a user whose request failed received no guidance at all. A one-line change raising the limit has been proposed to the production code but not applied.

**Earlier analyses.** A 16-case pilot gave 0% false reassurance for C, which looked like a complete fix. Zero of 16 has a 95% interval of 0 to 19%, so the pilot was too small to be informative, and the full hand-labeled result (44.9%) is far above it. An earlier 127-case build, which still contained the test products and used a looser metric, gave 15.2% for C against 17.7% for B with no significant difference ($p = 0.83$). After the dataset and the regex were corrected, the regex showed a significant B-versus-C difference (5.8% against 21.7%), but the hand labels do not (44.9% against 37.7%, $p = 0.40$), so the apparent disclosure effect on the regex measure was an artifact of the metric. Because data and metric changed together after we had seen results, we treat the present analysis as exploratory. We plan a confirmatory replication on newly collected cases with the metric definitions and analysis code frozen beforehand.

## VII. DISCUSSION

The clearest result is that an assistant without alert data sometimes describes conditions as calm or free of threat while an official alert is active (37.7% of cases under a neutral prompt), and that real alert data removed this in our replay (0 of 66 and 0 of 69 responses). The more useful result for builders is what the A2 condition shows: supplying the data was necessary to avoid false reassurance but not sufficient for safe behavior, because A2 raised alarms on quiet days that A did not. A preregistered single-sentence ablation (Section VI) confirms half of this: the one sentence telling the model how to treat the empty case does real, significant work when added to a sparse prompt, so instructions for the empty case are not optional. But it does not fully explain production's 0% false-alarm rate, and removing it from the full production prompt made almost no difference, so the practical lesson is narrower than it first looked: an explicit empty-case instruction helps substantially on its own, but a rich prompt's other framing can make the same sentence close to redundant once it is already present. Builders should not assume one sentence is either the whole story or beside the point; both halves of the ablation are informative.

Two design implications follow. First, an assistant should distinguish "no alert found" from "alerts not checked" and say which one applies. The unsupported-location state in the pipeline exists for this reason. Second, a disclosure instruction is not a substitute for the data. In our replay, telling the blind model to disclose its blind spot did not reduce false reassurance (44.9% against 37.7% without the instruction): the model disclosed in 12 of 69 responses and still described calm conditions or no immediate threat in 31. It also returned a non-"none" severity on every control, so it could not say that a quiet day is quiet. A cheap prompt-level fix for missing data should not be assumed to work without measuring it.

The test-product failure is a separate, practical lesson. A client that does not check the alert status field will show a user a test warning as real. We found ours only because the model happened to flag the alert as a test. The model's behaviour there was correct, but it is not a safeguard to rely on.

Finally, the study shows how easily small LLM evaluations mislead. A 16-case pilot suggested a prompt instruction fully fixed the problem, and an automated regex measure suggested it cut false reassurance from 21.7% to 5.8%, but hand labels showed no reduction at all. The regex caught 4 of the 57 false reassurances that hand labeling found (7.0% recall). The way to guard against this is to freeze definitions before collecting the data that will confirm them, and to check any automated metric against human labels before reporting it.

## VIII. LIMITATIONS

This is a pilot-scale study. The results should be read with these limits in mind.

- **Sample size.** 117 cases, of which 69 are alert cases and 59 are independent events. Several comparisons are significant, but confidence intervals are wide and per-hazard cells are mostly under 10 cases. The false-alarm results rest on only 32 controls.
- **Season and hazards.** Collection ran in late October. Tornado, tropical, heat and fire-weather alerts are essentially absent, and those are among the highest-stakes hazards. The results do not cover them.
- **Sampling bias.** The collection machine was asleep for 77% of the window, so short-lived warnings are likely under-represented. The bias cannot be corrected for this dataset.
- **Labeling and rule-based metrics.** False reassurance comes from hand labels by one rater, the author, who also built the system under test; there is no second rater and no reliability statistic. The coding scheme was written after an initial binary pass (Section V), and applying it moved the production pipeline's result from 9.1% to 0.0%. False alarm, hazard match and severity come from rules and keywords. The ablation conditions (A2plus and Aminus) were not hand-labeled, so their false-reassurance comparisons use the automated regex, which we found unreliable.
- **No advice-quality measure.** Advice correctness and harmful advice were not scored, so the study measures whether the assistant denied or noticed an alert, not whether its advice was right.
- **One pipeline, one model family.** The system, prompts and schema are from one project, and all calls used one model. Results may not transfer to other models or designs.
- **Headline-level alerts.** The production pipeline passes only headline-level alert fields. Passing the full alert text may change the results.
- **Metric and data changes.** The dataset and metric definitions were corrected after results were seen (Section VI), so the analysis is exploratory.
- **Partially isolated prompt effect.** A preregistered single-sentence ablation (Section VI) confirmed that one sentence in the production prompt substantially reduces false alarms when added to a neutral prompt, but did not fully close the gap to the production prompt's 0%, and removing the same sentence from the full production prompt made almost no difference. Some other part of production's prompt or schema is also doing work that has not been identified.
- **Failed calls.** Failed condition-A calls are excluded from the headline denominator; counting them as non-answers instead gives 0.0% false reassurance (0/69) and a 4.3% no-guidance rate (3/69), so exclusion does not flatter condition A on this measure, although users whose requests failed received no guidance.
- **Location.** Coordinates are city centres, not addresses, and a gridpoint alert can differ at street scale.
- **Unclassified alerts.** The NWS Special Weather Statement product cannot be classified from its event name alone and sits in an "other" family.
- **Author-built system.** The system under test was built by the author, which is a possible source of bias. The replay design, fixed seed and planned data release are meant to let others check the results.

## IX. CONCLUSION

An assistant that cannot see official alerts sometimes tells users that conditions are calm when an official alert is active. In a replay of 117 real cases, hand labeling of all 342 alert-case responses found this in 37.7% of responses under a blind neutral prompt (26/69, 95% CI 27.2–49.5) and 44.9% under a blind prompt that told the model to disclose its blind spot (31/69), and in none of the responses from the production pipeline (0/66) or from the same alerts under a neutral prompt (0/69). Every case was implicit; no response explicitly denied an alert. The difference between the blind conditions and those with alerts is large and significant (paired McNemar $p < 0.0001$), and the disclosure instruction did not help ($p = 0.40$). An automated regex measure suggested the opposite, with 21.7% and 5.8%, and caught only 4 of the 57 false reassurances that hand labeling found, so any such measure should be validated against human labels before it is used. Supplying alerts was necessary but not sufficient for safe behavior: with a neutral prompt, the model raised some alarm on 87.5% of quiet-day controls. A preregistered ablation showed that one sentence in the production prompt is a real but partial lever on that failure, substantial when added to a sparse prompt and close to redundant when removed from a rich one. We also found that test products can enter a naive pipeline as real alerts. These results come from 117 replayable cases in one late-October window, hand-labeled by a single rater who built the system, so they are a pilot. Next steps are a second rater with reliability statistics, scoring of advice quality, and a confirmatory replication over winter and spring hazards with frozen definitions.

## DATA AND CODE AVAILABILITY

We release the frozen case set, the case-construction and evaluation code, the prompts and schemas for all seven conditions (five main and two ablation), and the raw model responses, with a persistent DOI on Zenodo: https://doi.org/10.5281/zenodo.23901549. NWS data is published as open data for any use [10]. Open-Meteo data is licensed CC BY 4.0 for the free non-commercial API [11] and is attributed in the release.

## DECLARATIONS

- **Ethics.** The study used no human participants and no personal data.
- **Funding.** None. The model calls cost about $7 in total ($4.73 for the main study and $2.30 for the ablation) and were paid personally.
- **Competing interests.** The author developed the system under test (Stormwise), which was entered in the Amazon Developer Hackathon; if it wins, the author may receive prize money, with results expected in December 2026. The work was done on the author’s personal equipment at his own expense, with no funding from any organization, and the author declares no other competing interests.
- **Use of AI tools.** Per IEEE's policy on AI-generated content, two distinct uses are disclosed. (1) Claude Sonnet 4.6 (Anthropic), accessed via Amazon Bedrock, is the system under test in this study (Sections III, V and VI): its responses are the experimental data being evaluated, not content contributed to this manuscript. (2) Claude (Anthropic), run through Claude Code, assisted the author with writing and debugging the data-collection and evaluation code (Section V; the eval/ and logger/ scripts in the data release) and with drafting and revising manuscript text throughout, including the Abstract, Results, Discussion and Conclusion. In both uses, the author directed all analytical and editorial decisions, independently verified AI-assisted code against the underlying data and AI-assisted text against the computed results, and is responsible for the content of this article. No AI system is an author of this work.